\documentstyle[aps,prl,epsf]{revtex}

\def\beq{\begin{equation}}
\def\eeq{\end{equation}}
\def\bea{\begin{eqnarray}}
\def\eea{\end{eqnarray}}
\newcommand{\corr}[1]{\langle #1 \rangle}
\newcommand{\const}{{\rm const}\,}
\newcommand{\Tr}{\,{\rm Tr}\,}
\newcommand{\cT}{{\cal T}}

\begin{document}

\draft

\title{Supersymmetric Model of a 2D Long-Range Bose Liquid}
\author{M. V. Feigel'man and M. A. Skvortsov}
\address{L. D. Landau Institute for Theoretical Physics, Moscow 117940, RUSSIA}
\date{\today}
\maketitle

\begin{abstract}
The model Hamiltonian of a two-dimensional Bose liquid
(proposed earlier by Kane, Kivelson, Lee and Zhang \cite{kklz}
as the Hamiltonian
which has Jastrow-type wavefunctions
$\Psi_0\{{\bf r}_i\} \propto \prod_{j>k}|{\bf r}_j -{\bf r}_k|^{2\alpha}$
as the ground-state solution),
is shown to possess nonrelativistic supersymmetry. For the special
value of the coupling constant $\alpha =1/2$ the quantum mechanics described
by this Hamiltonian is shown to be equivalent
to the dynamics of (complex) eigenvalues
of random Gaussian ensemble of normal complex matrices.
For general $\alpha$, an exact relation between the equal-time current-current
and density-density correlation functions is obtained, and used to derive an
asymptotically exact (at low wavevectors $q$) spectrum of single-particle
excitations beyond the superfluid ground-state (realized at low $\alpha$'s).
The ground-state $\Psi_0$ at very large $\alpha$ is shown
to be of ``Quantum Hexatic" type, possessing long-range orientational order
and quasi-long-range translational order {\it but with zero shear modulus}.
Possible scenaria of the ground-state phase transitions  as function of
$\alpha$ are discussed.
\end{abstract}

\section{Introduction}

It is commonly believed that the system of Bose particles at zero temperature
may exist only in one of two possible ground-states:  superfluid  (SF) or
crystalline (CR).  Most probably it is indeed the case if the interaction
between bosons is short-range (compared to the mean interparticle distance
$n^{-1/2}$).  However the situation is much less clear if long-range
and/or long-time interactions are involved.  Two physical situations
where the problem of a two-dimensional (2D) Bose system interacting with
a 2D Maxwell gauge field were considered in \cite{fgil}: classical
 thermodynamics
of Vortex Line Liquid in High-Temperature Superconductors (see \cite{review}
for an extensive review) and low-temperature quantum properties of liquid
of Holons (arising in the RVB-type theories \cite{A} of doped Mott
dielectric \cite{IL}).  The arguments were given \cite{fgil} in favor
of existence of a new unusual ground-state which is still {\it liquid}, but
is {\it not superfluid}.  More recently the same problem was attacked
by another approach \cite{baskaran}, also in relation with RVB-type theories.
Still the theoretical arguments in favor of the existence of such unusual
ground-state are not quite conclusive; the results of different computer
simulations are in mutual disagreement \cite{MC1,MC2}. Whereas in \cite{MC1}
two phase transitions and an intermediate state were observed, in \cite{MC2}
a direct first-order transition from SF to CR state was found; note however,
that the model Hamiltonians simulated in these two numerical studies
were different in that the retarded interactions were presented  only in
\cite{MC1}.

There are two different questions related
with this problem: i) can one invent such a ground-state in a self-consistent
way? and ii) is it realized in some model with a physically realistic
long-range interaction?
In the present paper we try to approach the answer to
the first of the above questions, using a special model of a 2D Bose liquid,
introduced originally by Kane, Kivelson, Lee and Zhang (KKLZ) \cite{kklz}.
The KKLZ model Hamiltonian is constructed in such a way that
its exact ground-state wavefunction is of
the Jastrow type, i.e. it is a product of 1-particle and 2-particle
dependent factors. With the special choice of the Jastrow function the
main (in the long-wavelength limit) term of the KKLZ Hamiltonian is given
by a pair-wise logarithmic interaction, i.e. it resembles a 2D static Coulomb
potential of the real problem discussed in \cite{fgil}. On the other hand,
special symmetry (in fact, a kind of a supersymmetry) of the KKLZ
Hamiltonian makes it possible to obtain a number of exact relations between
the equal-time $T=0$ correlation functions, which help to analyze the general
structure of the theory.

The paper is organized as follows: in Section II we derive the KKLZ Hamiltonian
from the condition that it is given by the zero-Fermion sector of
a supersymmetric nonrelativistic parity-conserving theory,
which resembles (in the long-wavelength limit) the system of 2D bosons
with logarithmic interaction. The coupling constant $\alpha$ of our theory
is proportional to the charge of 2D bosons, thus
the ground-state possesses superfluidity at low $\alpha$ .
 In Sec.~III we analyze a special case ($\alpha=1/2$) of
the obtained model and show that it is equivalent to
the quantum mechanics of a Gaussian Matrix Model with normal complex matrices
(in the same sense as quantum mechanics of Gaussian Hermitean Matrix Model
is equivalent to 1D free fermions \cite{bla}); in Sec.~IV we are back to the
general case and derive an
exact real-space relation between the equal-time current-current
 and density-density
correlation functions (extending the similar results found in
\cite{tsa} for the Calogero-Sutherland model to a 2D case).
 These relations are then used in Sec.~V in order to derive a
relation between momentum-space transverse current-current
correlation function and the static structure factor, and in Sec.~VI to
find an exact form of the
 low-momentum excitation spectrum $\omega(q)$ (which appears to be given
by the one-loop correction to the MFA).  Sec.~VII is devoted to the discussion
of the ``crystalline" ground-state of the KKLZ Hamiltonian; we show that
such a state has in fact zero shear modulus and thus should be considered
as a kind of a ``quantum hexatic" (QHX), in some analogy with the classical
hexatic known to exist as an intermediate phase within the
Halperin-Nelson picture of thermal 2D melting \cite{HN}.
In Section VIII we speculate on possible properties of the additional
 ground-state which may be sandwiched between the SF  and QHX states.
 Our conclusions are  presented in Sec.~IX.
Finally, Appendix is devoted to the description a model of classical
 stochastic dynamics
which is equivalent to our quantum-mechanical problem; we show how to use
this equivalence for efficient numerical simulations of some
quantum-mechanical expectation values.

\section{The KKLZ Hamiltonian: SUSY structure and general framework}

We will see now that the KKLZ Hamiltonian can be considered as
an example of many-body generalization of the supersymmetric quantum
mechanics introduced in \cite{witten,ft}.
We define a pair of the conjugated ``supercharge" operators $Q$ and $Q^+$:
\begin{equation}
Q = \sum_{j,\alpha} q_{j,\alpha}a^+_{j,\alpha};
\qquad Q^+ = \sum_{j,\alpha} q^+_{j,\alpha}a_{j,\alpha};
\label{Q}
\end{equation}
where Fermionic operators $a_{j,\alpha}$ and $a^+_{j,\alpha}$ corresponds
to each component of the $j$-th particle's coordinate vector  $r_{j,\alpha}$
(with $\alpha = 1,2$), and
\begin{eqnarray}
q_{j,\alpha} = -i\frac{\partial}{\partial r_{j,\alpha}} +
  i\cdot 2\alpha
    \biggl(
      \sum_{k\neq j} \frac{\partial \ln|{\bf r}_j-{\bf r}_k|}
        {\partial r_{j,\alpha}} - \pi n r_{j,\alpha}
    \biggr),
        \\
\label{q}
q^+_{j,\alpha} = -i\frac{\partial}{\partial r_{j,\alpha}} -
  i\cdot 2\alpha
    \biggl(
      \sum_{k\neq j} \frac{\partial \ln|{\bf r}_j-{\bf r}_k|}
        {\partial r_{j,\alpha}} - \pi n r_{j,\alpha}
    \biggr).
\nonumber
\end{eqnarray}
Note that the introduction of two Fermionic operators $a_{j1}, a_{j2}$
is necessary
in order to construct a P-invariant theory; if the condition of P-invariance
were relaxed, the use of just one (per each particle)
Fermionic operator $c_j$ corresponding to the
holomorfic coordinate $z_j = x_j + i y_j$ would be possible.
Now we define the Hamiltonian
\beq
  {\cal H} = \frac{1}{2m} \{Q,Q^+\}
\eeq
and note that this Hamiltonian, when
acting on the state of ``Fermi-vacuum"
(defined by $a_{j,\alpha}|0_F\rangle = 0$),
$H={\cal H}|0_F\rangle$,
is equivalent to the Kane-Kivelson-Lee-Zhang (KKLZ) Hamiltonian \cite{kklz}
(see also \cite{fgil})
\beq
\label{kklz}
H - E^0 = H_{\rm KKLZ} = \frac{1}{2m}\sum_i
  \biggl\{
    - \nabla^2_i
    + 2\alpha
        \sum_{j\neq i} \nabla^2_i \ln|{\bf r}_i-{\bf r}_j| +
    4\alpha^2
        \sum_{j\neq i,k\neq i} \nabla_i \ln|{\bf r}_i-{\bf r}_j|
                               \nabla_i \ln|{\bf r}_i-{\bf r}_k|
        + (2\pi\alpha n)^2 {\bf r}_i^2
  \biggr\},
\eeq
where $E^0 = - 2\pi\alpha n N [1+\alpha (N-1)]/m$. The corresponding
ground-state wavefunction is given by
\beq
  \Psi_0({\bf r}_1,...,{\bf r}_N) = \const\!\cdot
  \prod_{j>k}|{\bf r}_j-{\bf r}_k|^{2\alpha}
  e^{-\pi\alpha n \sum_i r_i^2}.
\label{psi}
\eeq
Alternatively, the Hamiltonian can be written directly in terms
of the $q_{j,\alpha}$ operators:
\beq
H = \frac{1}{2m}\sum_j q^+_{j,\alpha}q_{j,\alpha}.
\label{kklz2}
\eeq
Each of the $q_{j,\alpha}$ operators annihilates the ground state (\ref{psi}),
so the ground-state energy corresponding to $H$ is zero.
The Hamiltonian (\ref{kklz}) contains 2-body and 3-body interactions;
in terms of the local particle's density $\rho({\bf r})$
 the corresponding potential energy is given by
\beq
U = \frac{1}{2} \int d^2r_1d^2r_2 \rho({\bf r}_1)\rho({\bf r}_2) V_2({\bf r}_1-{\bf r}_2)+
\int d^2r_1 d^2r_2 d^2 r_3
\rho({\bf r}_1)\rho({\bf r}_2)\rho({\bf r}_3)
V_3({\bf r}_1 - {\bf r}_2,{\bf r}_1-{\bf r}_3),
\label{U}
\eeq
where
$$
V_2({\bf r}) = \frac{4\pi\alpha}{m}\delta({\bf r}) +
\frac{4\alpha^2}{m} \frac{1}{r^2}; \qquad
V_3 ({\bf r},{\bf r}') =  \frac{2\alpha^2}{m}\frac{{\bf r\cdot r'}}{r^2r'^2}.
$$
Writing $\rho({\bf r}_1) = n + \delta\rho({\bf r}_1)$ where $n$ is the mean
density,  and integrating over ${\bf r}_1$ in the three-body term of
Eq.~(\ref{U}) we an find effective two-body potential
\beq
V_2^{\rm ef{}f}(r) = - \frac{g^2}{2\pi}\ln r + V_2(r),
\label{Ueff}
\eeq
where we introduced the Coulomb coupling  constant $g = 4\pi\alpha \sqrt{n/m}$
in the same way as in \cite{fgil}. Now the potential energy can be rewritten
in terms of the density variations $\delta\rho =\rho -n$, with the replacements
$\rho({\bf r}) \to \delta\rho({\bf r}) $ and $V_2({\bf r}) \to
V_2^{\rm ef{}f}({\bf r})$ made in Eq.~(\ref{U}).  On large space scales the main
term in the potential energy is the one containing $\ln{r}$, which leads to
the existence of a plasmon gap in the longitudinal branch of the excitation
 spectrum: $\omega(q = 0) = \omega_0 = \sqrt{ng^2/m}$.

Below we will be interested in the presence of superfluidity in the ground
state defined in Eq.~(\ref{psi}). We define superfluid density $n_s$ via
the variation of the ground-state energy $E_0$ (do not be confused by
the different quantities $E_0$ and $E^0$ in Eq.~(\ref{kklz}))
with respect to the external
(source) gauge field ${\bf a}$ minimally coupled to the
Hamiltonian (\ref{kklz}) via the replacement $\nabla_j \to \nabla_j +
i{\bf a}({\bf r}_j)$ in the kinetic energy term only:
\beq
  \frac{n_s}{m} = \left.
    \frac{\partial^2 E_0\{{\bf a}\}}
    {\partial {\bf a}_{\alpha}({\bf q})
    \partial {\bf a}_{\beta}(-{\bf q})} \right|_{q \to 0} \cdot
(\delta_{\alpha\beta}-\hat{q}_\alpha\hat{q}_\beta) = \frac{n}{m} -
{\cal P}(q\to 0,\omega=0),
\label{ns}
\eeq
where  the polarization function ${\cal P}(q,\omega=0)$ is related to
the transverse current-current correlation function
$\Pi_{\perp}(q,\omega)$
by the zero-$T$ FDT relation:
\beq
{\cal P}(q,\omega=0) = \frac{2}{\pi}\int_0^{\infty}
\frac{\Pi_{\perp}(q,\omega)}{\omega} d\omega.
\label{FDT}
\eeq
Below in Sec.~IV we will find the relation between
the {\it instantaneous} transverse current-current correlation function
$\Pi_{\perp}(q) = \frac{1}{\pi}\int_0^{\infty}\Pi_{\perp}(q,\omega)d\omega$
and the density-density correlation function.
Note that the replacement $\nabla \to \nabla + i{\bf a}$
 made in all terms in Eq.~(\ref{kklz}) would
retain SUSY structure of the theory and therefore would not change the
ground-state energy $E_0=0$.

The ground-state wavefunction is given by Eq.~(\ref{psi}) for any
value of the coupling constant $\alpha$, however this multiparticle
 wavefunction describes at least two ground-states which are very
different physically. Namely, at $\alpha \leq 1$
 the ground-state is a superfluid \cite{kklz} and its superfluid density
$n_s$ is equal to the total particle's density $n$.
However, superfluidity is not guaranteed by the form of the ground-state
wavefunction only:
at very large values of $\alpha$ (larger than $\alpha_{\rm cr} \approx 35$)
the ground state is expected to possess
translational symmetry breaking realized in the formation of a triangular
lattice \cite{wigner} (in fact only quasi-long Berezinsky-type
\cite{Berezinsky} translational order is formed). Such a state does not
possess superfluidity, $n_s = 0$.
Therefore at least one quantum phase
transition, between the states with nonzero and zero $n_s$, should
happen as the parameter $\alpha$ increases.  There is also a possible
room for the second phase transition. To see it, note that the Galilei
invariance of the Hamiltonian (\ref{kklz}) implies (at $T=0$)
 the absence of a normal density: $n_n = n - n_s = 0$. It means that
nonzero $n_n$ can appear due to spontaneous breakdown of the Galilei
invariance, which may happen simultaneously with the lattice formation
(so that $n_s$ jumps from $n$ to zero), or at some other critical value
$\alpha_G < \alpha_{\rm cr}$. Moreover, there are no special reason for
the superfluid density to go to zero at $\alpha = \alpha_{\rm cr}$, as it may
happen at $\alpha \in [\alpha_G, \alpha_{\rm cr}]$.  Finally, there is one
more symmetry, the orientational one, which can be spontaneously broken
by the ground-state at $\alpha = \alpha_O \in [\alpha_G,\alpha_{\rm cr}]$.
Below we start our consideration from the small-$\alpha$ range where no
symmetry is broken, then proceed (in Sec.~VII) to the
region of largest $\alpha > \alpha_{\rm cr}$, and at the end we
present some speculations on the system's behavior in the
intermediate range of $\alpha$.

\section{The special case $\alpha=1/2$: mapping to quantum mechanics of normal
matrices}

We will show now that  the dynamics of a Gaussian ensemble of  $N\times N$
complex matrices ($N\to\infty$)
constrained by the condition $[M,M^+]=0$ (which is the definition of a normal
 matrix)
is intimately related to the many-body quantum mechanics defined by the
Hamiltonian (\ref{kklz}) with $\alpha =1/2$.
This relation is of the same origin
as the well-known now relation \cite{bla} between dynamics of the Dyson
matrix ensembles and the Calogero-Sutherland model of many-body quantum
mechanics in 1 dimension, where the coordinates of 1D particles are
just eigenvalues of the corresponding matrices.
Eigenvalues of a normal matrix are,
in general,  complex numbers, which can be considered as coordinates of
particles living in a 2D space.  Thus our goal here is to show how it is
possible to perform functional integral over most part
of variables characterizing
dynamics of a normal matrix and reduce the problem to the quantum mechanics
for the set of its  complex eigenvalues $z_i, \bar z_i$.
Our derivation follows
closely  the one presented in \cite{bla} (see also \cite{gross}) for
three Dyson ensembles.  Consider the partition function
\beq
Z = \int {\cal D}\hat M(\tau) {\cal D}\hat M^+(\tau)\exp[-S(\hat M,\hat M^+)]
\prod_{\tau}\delta([\hat M(\tau),\hat M^+(\tau)])
\label{Z}
\eeq
with the flat measure of integration and the Gaussian action
\beq
S(\hat M,\hat M^+) = \int
  \left[\frac{1}{4}
    \Tr \biggl( \frac{\partial\hat M^+}{\partial\tau}
               \frac{\partial\hat M}{\partial\tau} \biggr) +
    \Tr (\hat M^+ \hat M)
  \right] d\tau.
\label{S}
\eeq
Note now that any normal matrix can be diagonalized by an appropriate
unitary transformation:  $\hat M = \hat U^+\hat Z\hat U$, where $\hat Z$ is diagonal
with the elements $z_i$.
Therefore we can change variables in the functional integral (\ref{Z}):
\beq
  {\cal D}\hat M(\tau) {\cal D}\hat M^+(\tau) =   J\{z_i\}
  \prod_i dz_i d\bar z_i {\cal D}\hat U,
\eeq
where  the Jacobian of the transformation is given \cite{mehta}
by the square of the absolute value of the Vandermonde determinant:
$J\{z_i\} =  \const |\Delta\{z_i\}|^2 = \const \prod_{i>j}|z_i-z_j|^2$.
Consider now the Hamiltonian corresponding to the above functional integral
in the transformed variables:
\beq
  H = - \frac{2}{\sqrt{g}} \sum_i
  \left[\frac{\partial}{\partial z_i}\sqrt{g}\frac{\partial}{\partial \bar z_i} +
  \frac{\partial}{\partial \bar z_i}\sqrt{g}
  \frac{\partial}{\partial z_i} \right]+
  \sum_i|z_i|^2 + \tilde H.
\label{Hmat}
\eeq
Here the first sum over $i$ constitutes a Laplacian operator in the
curved space with the square root of the determinant of the metric tensor
$\sqrt{g} = |\Delta\{z_i\}|^2 $, the second one comes from the
``potential energy" $\Tr (\hat M^+ \hat M)$, and the last term
corresponds to rotations of angular (i.e. related to unitary transformations)
variables, and thus vanishes when acting on any singlet ($\hat U$-invariant)
state. The states with nonzero angular momenta have much higher energy
\cite{grosskleb} and thus are irrelevant, therefore we will omit $\tilde H$ below.
Let us note now that the rest of the Hamiltonian (\ref{Hmat}) can be
rewritten in the form
\beq
  - \frac{4}{|\Delta\{z_i\}|}\sum_i
  \frac{\partial^2}{\partial z_i\partial\bar z_i} |\Delta\{z_i\}| +
  V\{z_i,\bar z_i\},
\label{Hmat2}
\eeq
where the potential energy $V$ is given by
\beq
  V\{z_i,\bar z_i\} = \sum_i\sum_{j\neq i,k\neq i}
  \frac{1}{(z_i-z_k)(\bar z_i-\bar z_j)}
  + 4 \sum_{i\neq j} \frac{\partial^2}{\partial z_i\partial\bar z_i}\log|z_i-z_j| +
  \sum_i |z_i|^2.
\label{V}
\eeq
After redefinition of the wavefunction $\Psi \to |\Delta\{z_i\}|\Psi$
and rewriting $z_i$ and $\bar z_i$ via real 2-vectors ${\bf r}_i$, we
are back to the Hamiltonian defined in Eq.~(\ref{kklz})
(apart from the constant term $E^0$) with $\alpha=1/2$,
$m=1/2$ and $n=1/\pi$.  Evidently, the values of $m$ and $n$ give just
the units of time and length
in the problem, and can be rescaled arbitrary. The  equilibrium probability
distribution  for the eigenvalues
(for the density of eigenvalues equal to $n$) is:
\beq
  P\{z_i\} = |\Psi_0\{z_i\}|^2 = \const\!\cdot \prod_{j>k}
  |z_j-z_k|^2
  e^{- \pi n\sum_i r_i^2}
\label{Pmat}
\eeq
and coincides (up to a constant normalizing factor) with the corresponding
distribution for the general complex ensemble \cite{ginibre,mehta} (cf.~also
recent preprints \cite{khare,oas}). Unfortunately it does not seem
possible to establish an exact relation between the
dynamics of general complex matrices and  some
Hamiltonian problem defined in terms of eigenvalues only;
the problem is that a general complex
matrix cannot be diagonalized by a unitary transformation and thus angular
variables cannot be eliminated from the dynamical problem \cite{note}.
It is only the stationary distribution of its eigenvalues
$P_{\rm complex}\{z_i\}$ which coincides (up to
a constant factor) with such a distribution for a normal matrix ensemble given
by Eq.~(\ref{Pmat}).
However, the relations obtained above indicate
that the correspondence between the KKLZ Hamiltonian and
normal matrices  is much deeper than just on the level of the
stationary distribution.
The supersymmetric structure of this Hamiltonian and the obtained relation with
a relatively simple matrix model lead us to believe
that the model under discussion
might be a rather direct generalization of the Calogero-Sutherland model for a
2D case, and thus might even be amenable for the exact solution.
However presently we are not aware of any exact way to treat normal
matrix dynamics, and therefore will turn in the next Sections
to some more elementary
methods, which still make it possible to find some exact relations between the
equal-time correlation functions of the KKLZ model at arbitrary value of the
parameter $\alpha$.

\section{The relation between the current-current and density-density correlation functions}

The SUSY structure of the Hamiltonian
(\ref{kklz}) makes it possible to find, using the method used previously
by Taniguchi, Shastry and Altshuler for the
Calogero-Sutherland model \cite{tsa} (see also recent preprint \cite{forr}),
an exact relation between the regular parts of the equal-time current-current
$\Pi_{\alpha\beta}^{\rm reg}({\bf r})$
and density-density
$K(r)$ correlation functions defined as
\beq
  \Pi_{\alpha\beta}^{\rm reg}({\bf r}-{\bf r'})
    = \corr{ J_{\alpha}({\bf r}) J_{\beta}({\bf r'}) }^{\rm reg}; \quad
  K({\bf r}-{\bf r'})
    = \corr{ \rho({\bf r})\rho({\bf r'}) }^{\rm reg}.
\label{Pi-K}
\eeq
In the definitions (\ref{Pi-K}) the singular (i.e.~proportional to
$\delta$-functions) terms in the current and density correlators are excluded
(i.e. ${\bf r} \neq {\bf r'})$. It means, in particular, that the
correlator $K({\bf r})$ is related to the inverse Fourrier-transform $s({\bf r})$ of
the static structure factor $S({\bf p}) = \int (d\omega) S(\omega,{\bf p})$
as follows:
\beq
  K({\bf r}-{\bf r'})
    = \corr{ 0|\sum_{j\neq k}\delta ({\bf r}-{\bf r}_j)
                             \delta ({\bf r'}-{\bf r}_k) |0 }
    = s({\bf r-r'}) - n\delta ({\bf r-r'}) + n^2,
\label{KS}
\eeq
where $n$ is the mean density of particles; the last term in (\ref{KS}) is due
to the fact that $K({\bf r})$ (contrary to $s({\bf r})$) is the reducible
correlation function). Similarly, the relation between the full current-current
correlation function and its regular part is:
\bea
  \Pi_{\alpha\beta}^{\rm reg}({\bf r}-{\bf r'})
    &=&
  \frac{(-i)^2}{4m^2}
  \corr{
    0|\sum_{j\neq k}
      \left(
        \frac{\partial}{\partial r_{j,\alpha}} \delta ({\bf r}-{\bf r}_j)
        + \delta ({\bf r}-{\bf r}_j) \frac{\partial}{\partial r_{j,\alpha}}
      \right)
      \left(
        \frac{\partial}{\partial r_{k,\beta}} \delta ({\bf r'}-{\bf r}_k)
        + \delta ({\bf r'}-{\bf r}_k) \frac{\partial}{\partial r_{k,\beta}}
      \right)|0
   } \nonumber
  \\
    &=&
  \Pi_{\alpha\beta}({\bf r}-{\bf r'}) -
     \frac{n}{m}E_{\rm kin}\delta_{\alpha\beta}\delta ({\bf r}-{\bf r}'),
\label{def}
\eea
where $E_{\rm kin} = \frac{m}{2N}\sum_i \corr{0|v_i^2|0}$
is the mean ground-state kinetic energy per particle
(in $D$ dimensions the last term in Eq.~(\ref{def}) should be
multiplied by $2/D$).

The trick to prove the relation between $\Pi_{\alpha\beta}^{\rm reg}({\bf r})$ and
$K({\bf r})$ is to substitute, following \cite{tsa},
the standard definition of the current operator
by the equivalent expression
\beq
  J_\alpha({\bf r})
    = \frac{1}{2m}\sum_i \left(q_{i,\alpha}^+\delta ({\bf r}-{\bf r}_i)
      + \delta ({\bf r}-{\bf r}_i) q_{i,\alpha}\right).
\label{Jnew}
\eeq
Then, using the commutation relations
\beq
  [ q_{i,\alpha}, q_{j,\beta}^+ ] =
    4\alpha
      \left[
        (1-\delta_{ij})D_{\alpha\beta}({\bf r}_i -{\bf r}_j) -
        \delta_{ij} \sum_{l\neq i}
        \left(
          D_{\alpha\beta}({\bf r}_i-{\bf r}_l)
          - \pi n \delta_{\alpha\beta}
        \right)
      \right]
\label{commut}
\eeq
with $D_{\alpha\beta}({\bf r}) = \nabla_\alpha\nabla_\beta \ln r$,
and identities $q_{i,\alpha}|0\rangle = 0$, we get
\beq
  \Pi_{\alpha\beta}^{\rm reg}({\bf r}) =
    \frac{\alpha}{m^2}\frac{\delta_{\alpha\beta}
    - 2 \hat{r}_\alpha \hat{r}_\beta}{r^2}
      K({\bf r}).
\label{PK}
\eeq
The relation (\ref{PK}) is applicable for any value
of the coupling constant $\alpha$ (remember that $\alpha \ll 1$ corresponds
to the superfluid ground state, whereas at very large
$\alpha \geq \alpha_{\rm cr} \approx 35$ the Wigner-type
crystal is formed \cite{wigner}).
One may wonder whether it is allowed to use the relation (\ref{PK}) at
$\alpha \geq \alpha_{\rm cr}$ since in course of its derivation
we have used space homogeneity of the
correlation functions, which  usually is not  present in the crystalline
state; however we will see below
that the ``crystalline" state present in our problem actually
does not possess a true positional long-range order and thus does not break
space homogeneity.

\section{Transverse currents and ``instantaneous" magnetic response}

We are going to obtain the momentum-space representations
for the longitudinal and transverse  current-current correlators defined as
$\Pi_{\alpha\beta}({\bf q}) =
\hat{q}_\alpha\hat{q}_\beta \Pi_{\parallel}({\bf q}) +
(\delta_{\alpha\beta}-\hat{q}_\alpha\hat{q}_\beta) \Pi_{\perp}({\bf q})$.
Performing the Fourrier transformation of the expression (\ref{PK})
and using the relation (\ref{def}), we get
\beq
  \Pi_{\alpha\beta}({\bf q}) =
  \frac{2\pi\alpha}{m^2}\int \frac{d^2p}{(2\pi)^2}
  \frac{(p+q)_\alpha(p+q)_\beta}{({\bf p+q})^2} {\tilde K}({\bf p})
  + \frac{2\pi\alpha}{m^2} n^2 \frac{q_\alpha q_\beta}{q^2}
  + \frac{n}{m}E_{\rm kin}\delta_{\alpha\beta}.
\label{fourrier}
\eeq
Here ${\tilde K}({\bf p}) = S({\bf p}) - n$ is the Fourrier-transform of
$K({\bf r})-n^2$.
In order to obtain the transverse polarization function
$\Pi_{\perp}({\bf q})$ one has to contract Eq.~(\ref{fourrier}) with the
transverse projector $\delta_{\alpha\beta}-\hat{q}_\alpha\hat{q}_\beta$:
\beq
  \Pi_{\perp}({\bf q}) =
  \frac{2\pi\alpha}{m^2}
  \int \frac{d^2p}{(2\pi)^2}
    \left[
      1 - \frac{[({\bf p+q}){\bf q}]^2}{({\bf p+q})^2 q^2}
    \right]
  {\tilde K}({\bf p})
  + \frac{n}{m}E_{\rm kin}.
\label{trans0}
\eeq
The $q\rightarrow 0$ limit of the frequency-dependent
$\Pi_{\perp}({\bf q}, \omega)$ is zero for any $\omega$, leading to the
same result for the equal-time correlator:
$\Pi_{\perp}({\bf q}\rightarrow 0) = 0$.
This condition makes it possible to determine $E_{\rm kin}$:
\beq
\label{kin}
  E_{\rm kin}
    = - \frac{\pi\alpha}{mn}  \int \frac{d^2p}{(2\pi)^2} {\tilde K}({\bf p})
    = - \frac{\pi\alpha}{mn} \left[ K(r=0) - n^2\right].
\label{ekin}
\eeq
Now, $K(r=0)=0$ since the ground state (\ref{psi}) forbids 2 particles
to be at the same point:
$\Psi_0(...,{\bf r},...,{\bf r},...)=0$.
Consequently, the expectation value of the kinetic energy in the ground
state is given by
\beq
  E_{\rm kin} = \frac{\pi\alpha n}{m} = \frac{\omega_0}{4},
\label{Ekin}
\eeq
where $\omega_0 = \sqrt{ng^2/m}$ is the plasma frequency. The relation
(\ref{Ekin}) was checked numerically (see Appendix for the details) and
found to be obeyed with a good accuracy for all values of $\alpha$, including
those there Quantum Hexatic phase is formed, where the Galilei and rotational
symmetries are both broken. We will discuss the implications of this
result below, in Sec.~VII.

Then, Eq.~(\ref{trans0}) may be rewritten as
\beq
  \Pi_{\perp}({\bf q})
    = - \frac{2\pi\alpha}{m^2}\int \frac{d^2p}{(2\pi)^2}
      \left[
        \frac{[({\bf p+q}){\bf q}]^2}{({\bf p+q})^2 q^2} - \frac12
      \right]
    \left(S({\bf p}) - n\right).
\label{trans}
\eeq

If the orientational symmetry is unbroken (which is the case for not
very large $\alpha$), the correlation function $S({\bf p})$
depends on the absolute value of the vector $\bf p$ only.
Then the angular integral in Eq.~(\ref{trans}) can be easily calculated,
the result is:
\beq
  \Pi_{\perp}(q) = - \frac{\alpha}{2m^2}\int_0^q (S(p)-n)
    \left( 1 - \frac{p^2}{q^2} \right) p\,dp.
\label{perp2}
\eeq
At small momenta $p$ the structure factor $S(p)$ goes to zero:
$S(p\to0) = np^2/2m\omega_0$, therefore
the low-momenta limit of the transverse correlator reduces to
\beq
\Pi_{\perp}(q\rightarrow 0) = \frac{\alpha n}{8m^2}\cdot q^2.
\label{chi}
\eeq
In the high-$q$ limit the integral in Eq.~(\ref{perp2}) reduces
to the one of Eq.~(\ref{ekin}):
\beq
\Pi_{\perp}(q\to\infty) = \frac{\pi\alpha n^2}{m^2}.
\label{piinfty}
\eeq
In the special case $\alpha =1/2$ the current-current correlation function
can be calculated explicitly.
Indeed, the real-space density-density
correlation function is known~\cite{ginibre,mehta}:
\beq
K_{1/2} (r) = n^2[1-\exp(- \pi nr^2)]
\label{Khalf}
\eeq
(this expression was derived by direct integration of the total
distribution function (\ref{Pmat}) over positions of all but two particles).
The corresponding structure factor is
\beq
S_{1/2} (p) = n [ 1 - \exp(-p^2/4\pi n)],
\label{s1/2}
\eeq
which leads, via integration in Eq.~(\ref{perp2}), to
\beq
\Pi_{\perp,1/2} = \frac{\pi n^2}{2m^2}
\left[1 - \frac{1-\exp(-q^2/4\pi n)}{q^2/4\pi n}\right].
\label{Pi1/2}
\eeq
Note that the current-current correlations in the Calogero-Sutherland
model were shown to be related to the correlations of eigenfunctions
of the corresponding random matrices \cite{tsa}; we expect that a similar
relation might exist between the correlation function (\ref{Pi1/2}) and
some eigenfunctions correlations in the random ensemble of normal
matrices.

We have employed numerical simulations (cf.~Appendix) to obtain
the structure factor $S(p)$ for several different values of
$\alpha$, and then used the obtained results
to compute, using Eq.~(\ref{perp2}), the transverse
current-current correlation function $\Pi_{\perp}(p)$. The results
are shown in Fig.~1.
The peaks at $p\sim 2\pi n^{1/2}$ correspond to a roton minimum in the
excitation spectrum that becomes deeper as $\alpha$ increases.
\begin{figure}[htb]
\epsfysize=4in
\centerline{\epsfbox{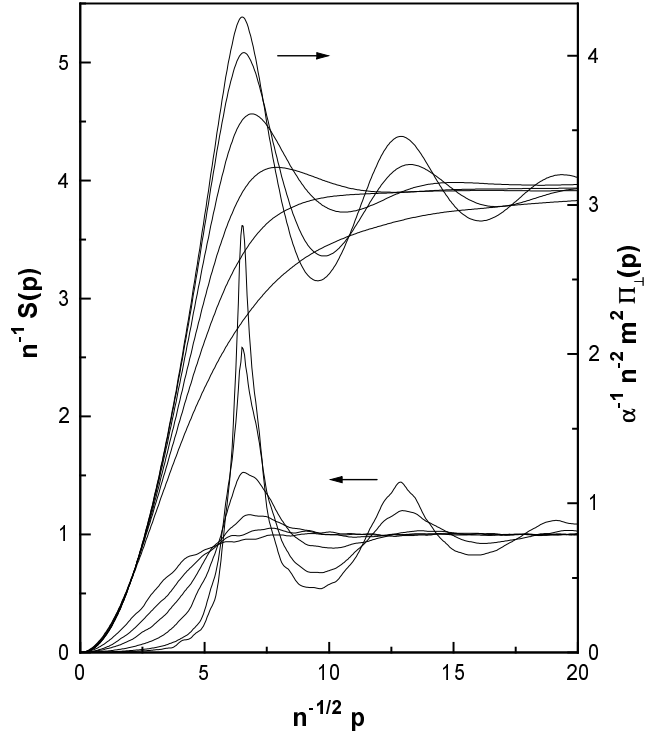}}
\caption{
The structure factor $S(p)$ (left axis) and
the transverse current-current correlator $\Pi_{\perp}(p)$ (right axis)
computed numerically by the mapping of the Schr\"odinger equation with
the Hamiltonian (\protect\ref{kklz}) onto a Langevin dynamics of classical
particles (see Appendix).
The curves correspond to $\alpha=$ 0.5, 1, 2, 5, 15 and 30.
The increase of $\alpha$ leads to the increase of the peak height at
$p\sim 2\pi n^{1/2}$.
}
\end{figure}

\section{Longitudinal currents and low-$q$ plasmon dispersion.}

Now we turn to the calculation of the longitudinal correlator
$\Pi_{\parallel}({\bf q})$.
Contracting Eq.~(\ref{fourrier}) with the longitudinal projector
and using (\ref{kin}), we obtain
\beq
  \Pi_{\parallel}({\bf q})
    = \frac{2\pi\alpha}{m^2}n^2
    - \Pi_{\perp}({\bf q})
    = \frac{2\pi\alpha}{m^2}n^2 +
    \frac{\alpha}{2m^2}\int_0^q (S(p)-n)
    \left( 1 - \frac{p^2}{q^2} \right) p\,dp.
\label{long}
\eeq

Alternatively, the value $\Pi_{\parallel}(0)$ may be found with
the help of the continuity equation that leads to the relation
$q^2\Pi_{\parallel}(q,\omega) = \omega^2 S(q,\omega)$
for the frequency-dependent correlators.
At $q \rightarrow 0$ the whole spectral weight is saturated by
the plasmon excitations with $\omega=\omega_0$, so we get
\beq
  q^2\Pi_{\parallel}(q)= \int \frac{d\omega}{2\pi} \omega^2 S(q,\omega)
  \stackrel{q\rightarrow 0}{\longrightarrow}
\frac{nq^2\omega_0}{2m}
\label{cont}
\eeq
leading to $\Pi_{\parallel}(0) = n\omega_0/2m$ in agreement with
(\ref{long}).

Eq.~(\ref{long}) can be used to find the low-$q$ dispersion of the
plasmon energy $\omega(q)$:
using the continuity equation
$q^2\Pi_{\parallel}(q,\omega) = \omega^2 S(q,\omega)$
together with the $f$-sum rule
$\int(d\omega)\omega S(q,\omega) = nq^2/2m$,
we convert Eq.~(\ref{long}) into the form
\beq
  \frac{\partial}{\partial q^2} \tilde\omega (q)
  = \frac{\alpha}{mn} \frac1{q^4}
    \int_0^q (S(p)-n) p^3\,dp,
\label{cont2}
\eeq
where
\beq
  \tilde\omega (q) =
  \frac{\int \omega^2 S(q,\omega) d\omega}{\int \omega S(q,\omega) d\omega}.
\eeq
As long as the spectrum $\omega(q)$ is nondecayable
[i.e.~${\rm Im}\,\omega(q)=0$], $\tilde\omega (q)$ coincides with
$\omega (q)$.

Finally, using the relation $S(q) = nq^2/2m\omega(q)$ valid at $q\to0$
and neglecting small $S(q)$ compared to $n$
in the r.h.s. of Eq.~(\ref{cont2}), we arrive at
\beq
  \tilde\omega (q) - \omega_0 = -\alpha \frac{q^2}{4m}.
\label{omega}
\eeq
Eq.~(\ref{omega}) is very similar to the one derived \cite{sf} in the
lowest-order perturbation theory (over $\alpha$)
for the purely Coulomb 2D Bose-liquid
(the result of \cite{sf} contains extra numerical factor $\approx 0.8$ in
the r.h.s.~of Eq.~(\ref{omega})). Let us emphasize, however, that in the
present model we are able to derive the relation (\ref{omega}) for
{\it any} value of $\alpha$, the only limitation was the low-$q$ limit.
In the field-theoretic language,
the low-momentum excitation spectrum of the KKLZ
Hamiltonian is given by the first-loop term in the diagram expansion.
We believe that this remarkable property is due to the supersymmetry of the
model, leading somehow to the cancellation of all higher-order diagrams.
It is not yet clear to us whether the one-loop calculation is exact
not only for the low-$q$ spectrum, but for the spectrum and the correlation
functions in general.
To study this issue one needs to develop an explicitly supersymmetric
second-quantized representation and the corresponding
diagram technique for the KKLZ model.

As long as $\alpha \leq 1$, the quadratic dispersion $\omega (q)$ given by
Eq.~(\ref{omega}) produces a relatively weak minimum between $\omega_0$ and
the free-particle limit $\omega_{\rm free}(q) = q^2/2m$.  Such a spectrum
is consistent with a totally superfluid ($n_s=n$) ground state.
With the increase of $\alpha$ the minimum in the spectrum becomes deeper;
for a qualitative description of the spectrum one can use
a kind of an interpolation formula
\beq
\omega^2(q) \approx  \omega_0^2 - \omega_0\alpha q^2/2m + (q^2/2m)^2.
\label{interpol}
\eeq
The minimum of the r.h.s. of Eq.~(\ref{interpol}) touches
 zero at $\alpha = 2$.  It can be considered as an indication of
possible spectrum instability and phase transition(s) in the large-$\alpha$
region. However the nature of Eq.~(\ref{interpol}) as interpolation formula
of unknown (at large $\alpha$) accuracy makes it impossible to use it for
any exact conclusion about the behavior of the spectrum. In the next Section
we will try to consider this problem starting from the opposite limit of
extremely large $\alpha$.

\section{``Crystalline" ground state: quantum hexatic without shear modulus}

At very large values of $\alpha > \alpha_{\rm cr} \approx 35$ the ground state is expected to possess
symmetry breaking realized in  the formation of triangular
lattice \cite{wigner}. The static structure factor $S(p)$ can be determined
knowing the ground state wavefunction only. Its calculation
involves integration over all particle's coordinates ${\bf r}_i$ with
the weight given by $\Psi_0^2\{{\bf r}_i\}$, which
is formally
equivalent to the calculation of the structure factor of a classical 2D
crystal with a logarithmic repulsion between particles
$V(r) = 4\alpha\cdot \ln r$
and ``temperature" ${\cal T} =1$ (cf.~Appendix).
Such a classical crystal has a finite shear
modulus $\mu_{\rm cl} \sim \alpha n$
and is expected to undergo dislocation-mediated
melting when $\alpha$ decreases below some critical value $\alpha_1$.
At $\alpha > \alpha_1$ it is stable with respect to dislocation formation,
but  does not possess long-range
translational order (as any other 2D crystal at finite temperature).
It means that the structure factor $S(p)$ does not have Bragg peaks near
the principal inverse lattice vectors ${\bf p}_a$, but shows only power-low
singularities:
\beq
  S(p) \sim |{\bf p} - {\bf p}_a|^{\eta-2}
\label{berez}
\eeq
with $\eta \sim \alpha^{-1}$.
Let us now go back to our original problem, which is a
quantum 2D crystal. Suppose that it has finite shear modulus $\mu$ (now without
subscript `cl'), then the spectrum of the transverse sound would be linear,
as usual: $\omega(q) \sim \sqrt{\mu}q$ where $q$ is the wavevector of a shear
deformation
on a top of the crystal state. (In terms of momenta ${\bf p}$
of the original Bose particles it corresponds to
$\omega_{\perp}({\bf p}) \sim |{\bf p} - {\bf p}_a| $ for any $a$).
Now we can calculate the zero-point fluctuations of lattice displacements
$\corr{u^2} \sim \int d^2q \omega^{-1}(q)$ and find it finite, which means a
finite Debye-Waller factor and thus the presence of Bragg peaks in the
structure factor $S(p)$ -- in contradiction with the above-obtained
Eq.~(\ref{berez}).
The only way out of this contradiction is to suppose that the quantum
solid with our very special interaction given by Eq.~(\ref{kklz}) has
actually {\it zero} shear modulus: $\mu = 0$. More precisely, we need to
suppose that $\omega (q) \propto q^2$, which leads then to a logarithmically
divergent $\corr{u^2}$, in agreement with Eq.~(\ref{berez}). It means that the
``crystalline" ground state expected to exist at $\alpha > \alpha_1$ should
be more adequately called ``Quantum Hexatic" (QHX), which is (in close
 analogy to the classical hexatic phase introduced by Halperin and Nelson
within their picture of dislocation-mediated thermal melting of 2D
crystals \cite{HN}) the phase without shear modulus
but still with anisotropy of the excitation spectrum:
\beq
  \omega_{\perp}({\bf p}) \sim ({\bf p} - {\bf p}_a)^2/M
\label{hex}
\eeq
($M$ is some parameter with dimensionality of mass),
whereas in the isotropic liquid excitation spectrum depends on the
momentum absolute value $|{\bf p}|$ only.

The absence of a shear modulus in the high-$\alpha$ ground-state of
the Hamiltonian (\ref{kklz}) can also be shown by direct calculation
which can be conveniently performed representing potential
energy of interaction between particles as
\beq
  U_{\rm KKLZ} = \frac{2\alpha^2}{m}\sum_i
  \biggl(
    \sum_{j\neq i}\frac{{\bf r}_j-{\bf r}_i}{({\bf r}_j-{\bf r}_i)^2}
  \biggr)^2
  + \frac{2\pi\alpha}{m}\sum_{i\neq j}\delta({\bf r}_i-{\bf r}_j)
 \label{poten2}
\eeq
Now one can calculate the variation of the potential energy (actually
the second sum in (\ref{poten2}) does not contribute)
corresponding to an infinitesimal uniform shear deformation
$\epsilon_{\alpha\beta}$ of the regular lattice. Then one finds that it
does not contain the term proportional to $\epsilon_{\alpha\beta}^2$, i.e.
the expansion starts from the term
$(\nabla \epsilon_{\alpha\beta})^2$, which means that $\mu (q) \propto q^2$.

The absence of a genuine crystalline state in the KKLZ model at arbitrary
strong interaction $\alpha \to \infty$ stems from its very special form
and certainly is not present in a generic 2D Bose system.
 However, in a more realistic system  QHX may exist
as an intermediate ground-state realized at moderately large repulsion.

Let us consider now the response of the QHX state to a magnetic
field $h$.  Since particles in this state are almost localized, we can
neglect (in a first approximation) the multi-particle exchange contributions
and estimate susceptibility as
$\chi \sim \frac{n}{m} \corr{u^2} \sim \frac{M}{m} \ln(nL^2)$ (where $L$ is the system's
size).  Comparing this expression with the magnetic susceptibility
of a superfluid, which is proportional to $n_s L^2$, one concludes
that $n_s=0$ in the QHX state.

Now we need to address some apparent paradox: i) according to
Eq.~(\ref{FDT}), normal density $n_n = n - n_s$ is proportional
to the integral
$\int_0^{\infty}\Pi_{\perp}(q\to 0,\omega)d\omega/\omega$ where the expression
under the integral is positive; ii) the equal-time current-current
correlation function
$\Pi_{\perp}(q) = \int_0^{\infty}\Pi_{\perp}(q,\omega)d\omega/\pi$
vanishes at $q\to 0$ (cf.~Eq.~(\ref{Ekin}));
 iii) then, how it is possible to get nonzero $n_n$
(and even cancel $n_s$ completely)?
 In fact, to get nonzero $n_n$ we need to allow for the
singular behavior of the function
$\Pi_{\perp}(q,\omega)$ at low $q,\omega$ (in other words, the limiting
procedure $q\to 0$ and integration in Eq.~(\ref{FDT}) should not commute
with each other). As a simple example one can consider
$\Pi(q,\omega) \propto f(q^2/m\omega)$ with $f(x) = x/(1+x^2)$.
However, such a behavior of $\Pi(q,\omega)$ is only possible if there are
some gapless excitations in the system, which is just the case in the
QHX state (cf.~Eq.~(\ref{hex})).
In general,
an appearance of gapless excitations with nonzero momentum (low-momenta excitations
in our system always have large plasmon gap $\omega_0$) may only be a result
of a phase transition, which is associated with spontaneous breakdown
of the Galilei invariance. Generally speaking, the Galilei invariance may
be spontaneously broken  at $\alpha = \alpha_G < \alpha_{\rm cr}$, which would
mean the existence of two phase transitions and some intermediate state
between the SF and the QHX states.

\section{Between Superfluidity and Quantum Hexatic: Speculations}

We found in previous Sections that at small $\alpha$ the ground-state
$\Psi_0$ possesses total superfluidity ($n_s=n$) and the excitation spectrum
beyond the ground state is gapful (cf.~Eq.~(\ref{interpol})).
At very large $\alpha \geq \alpha_{\rm cr} \approx 35$ the ground state is
not superfluid at all, $n_s=0$, and the excitation spectrum is very soft
(cf.~Eq.~(\ref{hex})). How does the transition(s) between these two
qualitatively different ground states occur? The most trivial possibility
would be that of the direct first-order phase transition at
 $\alpha = \alpha_{\rm cr}$. This would also mean that the first-order
melting/crystallization transition happens
in the classical 2D Coulomb gas at the interaction parameter
$\Gamma = 4\alpha \approx 140$. However, the persistence of total
superfluidity up to $\alpha \approx 35$ looks rather improbable
since at so large values of $\alpha$ the characteristic energy of the
interaction between particles, $g^2/2\pi$, exceeds the
kinetic energy $E_{\rm kin} = \pi\alpha n/m$
by more than 2 orders of magnitude.
Note also, that the above-cited value of $E_{\rm kin}$ is about 100 times
larger than the naive estimate $n/m$; it means that particles are almost
localized in rather small regions of space (much smaller than the mean
interparticle distance $n^{-1/2}$) and diffuse from these regions on
much longer time scale only.
Therefore we suspect that at least two phase
transitions take place between the totally SF state and the QHX state.

As we discussed above, the QHX quantum ground-state is seen as a 2D
Berezinsky crystal with logarithmic interaction as long as
only the equal-time density correlations are considered.
The nature of melting in such a 2D classical crystal is not quite clear.
The old numerical
simulations reported weak first-order transition \cite{wigner}, but
the recent simulations on the 2D systems with other types of interactions
\cite{newMC} reveal that to discern between 1st and 2nd order melting one
probably would need to simulate systems with a huge number of particles,
which was not yet done for the system of interest here.

If melting of such a 2D crystal occurs via Halperin-Nelson mechanism,
then there exists a usual hexatic state
which possesses quasi-long-range
orientational order, i.e. power-law decay of the orientational order
parameter
\beq
  \langle \exp[6i(\theta(0)-\theta({\bf r}))]\rangle \propto r^{-\eta_6}
\label{6}
\eeq
without any translational order. The corresponding quantum
ground-state would be characterized, apart from orientational quasi-LRO,
by the isotropic structure factor $S(p)$ and by the existence of some gapless
mode related with a slow rotation of the $\theta ({\bf r})$ angle.
The other possibility is that the melting of the
2D crystal with log-interaction
occurs via a first-order phase transition, without hexatic phase;
even in this case there is a room for a phase transition related to
the breakdown of the Galilei invariance, which would be manifested by
an appearance of the gapless mode in the excitation spectrum.
 Indeed, the absence of normal density
$n_n$ in the ground-state of the system with purely potential interactions
is the consequence of the Galilei invariance of the Hamiltonian, which
is certainly respected in the low-$\alpha$ limit.  An appearance of non-zero
$n_n$ is thus related to spontaneous breakdown of the Galilei invariance,
which may occur at some value of $\alpha$ which we denote as $\alpha_G$.
As it was explained above, we expect $\alpha_G < \alpha_{\rm cr}$, although
we still cannot prove this inequality.

To conclude this Section: if there exists any additional ground-state
sandwiched between SF and QHX, then the spectrum of excitations
beyond such a state
contains a gapless mode with isotropic spectrum:
$\omega(p) \propto (p-p_0)^{\beta}$.  Such a mode may be orthogonal to
the density mode and, thus, it may be invisible in the usual structure
factor $S(q,\omega)$.

\section{Conclusions}

We have studied the model Hamiltonian
of the two-dimensional strongly interacting
Bose-liquid, which can be considered as a natural generalization of the
Calogero-Sutherland Hamiltonian for the 2D case. This Hamiltonian
is shown to possess the non-relativistic supersymmetry. It makes possible
to derive a number of exact relations
between the equal-time density-density and
current-current correlation functions, and to find exactly the low-momentum
excitation spectrum of the model, which is given by the one-loop correction
to the mean-field result.  The ground state is shown to exist in at least
two qualitatively different phases as function of the interaction parameter
$\alpha$: the
superfluid state at small $\alpha$ and the quantum hexatic state at very
large $\alpha$.
The arguments are presented in favor of the existence of
a third state, intermediate between the two above-mentioned ones.
For the special case of $\alpha = 1/2$ the considered model is shown
to be equivalent to the quantum mechanics of normal complex matrices with
a Gaussian weight, exactly in the same way as the Calogero-Sutherland model at
the special values of its coupling constant is
equivalent to the dynamics of the Wigner-Dyson matrix ensembles.

The model we have studied seems to be very interesting as a natural
approximation for the problem of 2D Bose-liquid with logarithmic pair-wise
interaction, which is widely considered in relation with the composite-boson
approach to the description of the Fractional Quantum Hall
state~\cite{kklz,dhlee},
as well as in relation with statistics of vortex lines
in HTSC superconductors~\cite{fgil,MC2}. The key physical problem here is
the problem of existence of  non-superfluid liquid ground-state. We have
shown within the considered model the existence of at least one nontrivial
ground-state, which is quantum hexatic, i.e. anisotropic non-superfluid
liquid. It is still an open question if the destruction of superfluidity
may occur without {\it any} explicit symmetry-breaking in the real space,
i.e. if it might be that the only spontaneously broken symmetry is the
 Galilei invariance.
To answer this question it is necessary to study the spectrum of
 lowest-energy excitations at large wavevectors $q\sim 2\pi\sqrt{n}$
and at large $\alpha \in [2,35]$ and look for the appearance of some
gapless mode; note that such a mode, if it exists, may be orthogonal to
the density-oscillation mode, and thus may be absent from the dynamic
structure factor $S(q,\omega)$.

The exact results obtained make it possible to suspect that the considered
model can actually be exactly solvable, at least at some special values
of $\alpha$. There are several directions in which one could try to extend
and develop the results presented here:
i) to find the time-dependent correlation functions,
at least for $\alpha =1/2$, by the solution of the normal-matrix problem;
ii) to generalize the correspondence between our Hamiltonian
and the normal matrix problem for other values of $\alpha$;
iii) to develop an explicitly supersymmetric second-quantized
many-body theory and the corresponding
diagram technique for our Hamiltonian; this might be very helpful in order
to generalize our result for the spectrum, Eq.~(\ref{omega}), for an
arbitrary momenta $q$.
A completely different direction of study of this model is provided
by its equivalence to a Fokker-Plank equation of multi-particle stochastic
dynamics (cf.~Appendix), which makes it possible to use {\it classical}
Monte-Carlo simulations in order to study some properties of the
quantum many-body problem in question.

We are grateful to V.B.Geshkenbein, A.Gorsky, A.I.Larkin, A.Yu.Morozov,
A.M.Polyakov and A.S.Schwartz for useful discussions.
We are especially indebted to L.B.Ioffe
for his collaboration on earlier stages of this research and many
useful advises.
The hospitality of the Isaac Newton Institute where this study
was initiated during the visit of one of us (M.V.F., August 1994) is
gratefully acknowledged. This research was supported by the
INTAS-RFBR grant \# 95-0302, Swiss National Science Foundation
 collaboration grant \# 7SUP J048531, and
DGA grant \# 94-1189 (M.V.F.).

\section*{Appendix: SUSY, Langevin dynamics and numerical simulations}

\begin{appendix}
The relation between supersymmetric quantum mechanics and Fokker-Planck
equation of stochastic classical dynamics is well-known now
\cite{ft,zinn} (cf.~also recent preprint \cite{forr}).  In this Appendix
we will show how to use that equivalence in order to implement fast
numerical simulation of the many-body quantum-mechanical problem we
are interested in.

Consider a system of $N$ coupled Langevin equations:
\beq
\label{Langevin}
  \gamma \frac{dr_{j,\alpha}}{dt} =
    - \frac{\partial W\{{\bf r}_i\}}{\partial r_{j,\alpha}}
    + \xi_{j,\alpha}(t)
\eeq
with a Gaussian distribution of random forces $\xi(t)$:
\beq
  \overline{ \xi_{j,\alpha}(t) \xi_{k,\beta}(t') } =
    2\gamma \cT \delta_{jk} \delta_{\alpha\beta} \delta(t-t').
\eeq
Here $\cT$ is temperature of a thermal bath,
$\gamma$ is a damping coefficient which can be set to 1 by a proper rescaling
of time.

The Fokker-Planck equation for the probability density function,
\beq
\label{FP}
  \gamma \frac{\partial {\cal P}\{{\bf r}_i\}}{\partial t} =
    \sum_{j,\alpha}
    \frac{\partial}{\partial r_{j,\alpha}}
      \left(
        \frac{\partial W}{\partial r_{j,\alpha}}
        + \cT \frac{\partial}{\partial r_{j,\alpha}}
      \right)
    {\cal P}\{{\bf r}_i\},
\eeq
after the substitution ${\cal P}\{{\bf r}_i\} = e^{-W/2\cT} \Psi\{{\bf r}_i\}$
reduces to an imaginary-time Shr\"odinger equation
\beq
\label{SE}
  \gamma \frac{\partial \Psi}{\partial t} =
    \sum_{j,\alpha}
    \left\{
      \cT \frac{\partial^2 \Psi}{\partial r_{j,\alpha}^2} +
      \left[
        \frac12 \frac{\partial^2 W}{\partial r_{j,\alpha}^2} -
        \frac{1}{4\cT}
          \left(
            \frac{\partial W}{\partial r_{j,\alpha}}
          \right)^2
      \right]
      \Psi
    \right\}.
\eeq
The many-particle ground state of Eq.~(\ref{SE}) has zero energy,
which is a consequence of a supersymmetry \cite{ft}, and
wavefunction $\Psi_0 = \const e^{-W\{{\bf r}\}/2\cT}$.

An equivalence of Langevin, Fokker-Planck and Shr\"odinger approaches
implies that the averaging over many-particle ground state $\Psi_0$
is equivalent to the averaging over the Langevin dynamics of particles
(with respect to random force and initial particles distribution)
\cite{forr}:
\beq
\label{equiv}
  \corr{\Psi_0|F\{{\bf r}_i\}|\Psi_0} =
  \overline{ \corr{F\{{\bf r}_i\}} }_L.
\eeq

The relation (\ref{equiv}) is very important as it makes possible to
compute ground-state averages by numerical simulation of Langevin dynamics
with a proper interparticle interaction $W$ which is technically
substantially simpler than direct quantum Monte-Carlo simulations.
The problem at hand, given by the wavefunction (\ref{psi}), corresponds
to a stochastical classical dynamics of a one-component 2D Coulomb gas
governed by the following potential energy:
\beq
\label{W}
  W\{{\bf r}_i\} =
    - 4\alpha\cT \sum_{j>k} \ln|{\bf r}_j-{\bf r}_k|
    + 2\pi\alpha\cT n \sum_i r_i^2.
\eeq

We have performed Monte-Carlo computer simulations of the Brownian dynamics
(\ref{Langevin}) with the potential (\ref{W}) for the systems of $N=109$
particles with different values of the parameter $\alpha$. The results
obtained for the structure factor $S(p)$ are shown in Fig.~1.
Also shown are the curves for the correlation function of transverse currents,
$\Pi_\perp(p)$, calculated from $S(p)$ according to Eq.~(\ref{perp2}).

We have also used this simulation in order to check Eq.~(\ref{Ekin})
for different values $\alpha$. Note that the general relation (\ref{equiv})
can not be applied to compute the mean value of the kinetic energy since
the operator of the kinetic energy contains space derivatives that have no
immediate meaning in the classical Langevin dynamics. Fortunately,
instead of $E_{\rm kin}$ one can compute the mean ground-state potential
energy $E_{\rm pot}$ using the fact that the ground-state energy is zero.
$E_{\rm pot}$ may be easily obtained
by Langevin simulations using Eq.~(\ref{equiv})
since the operator of the potential energy
$U_{\rm KKLZ}+E^0+\frac{(2\pi\alpha n)^2}{2m}\sum_i r_i^2$ contains
particles coordinates only.

Numerical simulation have showed that Eq.~(\ref{Ekin}) holds
for all values of $\alpha$ from $\leq 1$, that corresponds to the SF state,
to $100$, corresponding to the CR state,
within the precision of computation.
\end{appendix}

\end{document}